# Advance Reservation based DAG Application Scheduling Simulator for Grid Environment

Harshad B. Prajapati[1] and Vipul A. Shah[2]

[1] Information Technology Department,
Dharmsinh Desai University, Nadiad-387001, INDIA

[2] Instrumentation and Control Engg. Department,
Dharmsinh Desai University, Nadiad-387001, INDIA

*Abstract*--In the last decade, scheduling of Directed Acyclic Graph (DAG) application in the context of Grid environment has attracted attention of many researchers. However, deployment of Grid environment requires skills, efforts, budget, and time. Although various simulation toolkits or frameworks are available for simulating Grid environment, either they support different possible studies in Grid computing area or takes lot of efforts in molding them to make them suitable for scheduling of DAG application. In this paper, we describe design and implementation of GridSim based ready to use application scheduler for scheduling of DAG application in Grid environment. The proposed application scheduler supports supplying DAG application and configuration of Grid resources through GUI. We also describe implementation of Min-Min static scheduling algorithm for scheduling of DAG application to validate the proposed scheduler. As the proposed DAG application scheduler is based on Java language, it is extensible and portable. Our proposed DAG application scheduling simulator is useful, easy, and time-saver.

*Index Terms*—DAG application scheduler, DAG application scheduling, dependent task scheduler, dependent task scheduling, static scheduling, Grid simulation.

## I. INTRODUCTION

GRID computing [1] enables execution of performance demanding scientific-applications by exploiting resources in collaborative manner. In Grid, scheduling is performed at two levels by two different entities, which are called resource schedulers and application scheduler. The scheduling [2], [3], [4] aspect in Grid computing decides about which tasks would be executed on which resources; what would be the order of tasks; and how data dependencies would be respected. The main objective of application scheduling is to minimize the makespan of the Directed Acyclic Graph (DAG) application. As the nature of problem of DAG application scheduling is NP-Complete [5], research in scheduling of DAG application has resulted in many heuristics.

In Grid computing, a solution of a scientific problem can be implemented by composing existing available executables in form of a scientific workflow [6], [7], [8]. Such scientific workflow application is represented as DAG. The DAG application is scheduled by an application scheduler or an application broker. A real Grid environment can be used to carryout research in scheduling of DAG application. However, using real Grid environment for testing of newly proposed scheduling heuristics may not be timely and budgetary feasible for all the researchers, as this approach of real experimentation in Grid environment demands expert skills in deployment, a big budget for a real Grid test-bed, and longer time to wait for getting results. Moreover, enabling certain test scenarios in Grid test-bed becomes difficult and in some cases impossible. In such situations, the simulation based approach of problem solving and solution understanding becomes very useful.

Various simulators or simulation frameworks are available supporting simulation based study in Grid environment. However, they differ based on kind of study supported by them. For example, GridSim [9], SimGrid [10], ChicSim [11], OptorSim [12], SimBOINC [13], and Alea [14] are few of them. The GridSim and SimGrid toolkits deserve for discussion in this research paper in context of scheduling of DAG application in Grid environment. The GridSim toolkit supports time-shared, space-shared, and advance reservation space shared policies as resource scheduler policies. Moreover, for application scheduling, earlier version of GridSim supported scheduling of parametric applications. However, there is no direct support for scheduling of DAG application in GridSim. The second one, SimGrid has support for DAG API. However, this toolkit is a generalized toolkit for distributed computing environment; moreover, in SimGrid, there is no direct support for simulating Grid environment by creating various Grid entities such as cluster, task, Grid Information Service, Time-shared resource, space-shared resource, and many more, which are supported by GridSim. This is the reason behind research on our own simulator supporting scheduling of DAG Application in Grid Environment. Due to a rich set of classes and functions available in GridSim, we based our DAG Application Scheduling Simulator for Grid Environment on the GridSim toolkit.

Our major contributions in this paper are: (1) We propose the design of GridSim based architecture supporting simulation of DAG application scheduling algorithms, (2) We provide the implementation of the proposed architecture, (3) We validate our proposed, implemented scheduler by implementing min-min scheduling algorithm using our proposed application scheduler.

Our scheduling simulator allows a user to prepare task-graph using GUI by performing drag and drop. Moreover, while preparing a DAG, the user can specify the amount of computation needed by each task and communication amount between a pair of tasks. It also supports configuring resource information through GUI. These features save valuable time of researcher, as the researcher does not need to write code for creating tasks and Grid resources manually. Moreover, our DAG application scheduling simulator handles many mundane activities such as traversal of abstract DAG, (2) sorting list of resources as per their execution cost, (3) sorting list of resources as per their execution time. The proposed DAG application scheduling simulator is extensible, as it can allow testing of new scheduling algorithm. This proposed DAG application scheduling simulator would enable performance evaluation of existing algorithms, testing of new proposed algorithm before deploying it on real Grid environment, and usage in educational purposes.

This research paper is structured as follows. Section II describes the problem of DAG application scheduling and what support is available from GridSim for scheduling of DAG application. Section III presents the proposed DAG application scheduling simulator and discusses the important implementation details. Section IV shows the implementation of Min-Min scheduling algorithm in the proposed simulator and presents the result generated on our proposed simulator. Finally, Section V presents conclusion and possible future work.

## II. Problem Statement and Support from GridSim for DAG application scheduling

### A. DAG application scheduling problem

A DAG is a graph data structure having no cycle. It is used to represent nodes and dependency among the nodes. In Grid computing environment, the tasks are represented as nodes of DAG and data-communication between tasks is represented as an edge of DAG. An example of DAG is shown in Figure 1. This example DAG has two parallel tasks, node 2 and node 3, in addition to entry task, node 1, and exit task, node 4. The number besides each node represents amount of computation needed by the task, and the number on an edge between a pair of tasks represents the amount of data communication happening from a source task to the destination task of the edge. Formally, a DAG can be represented as G = (V, E), where V is set of vertices or nodes represented as $V_i$, where i = 1…|V|. The E is a set of vertices or edges representing precedence constraints between a pair of nodes or vertices. The E is set of $e_{i,j}$, where i identifies the source task and j identifies the destination task of the current dependency.

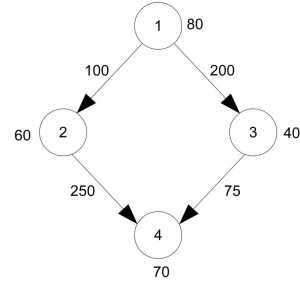

Figure 1  An example Directed Acyclic Graph representing tasks and dependencies among the tasks

While scheduling of DAG application is done, the sequential tasks need to be executed in order of sequence and parallel tasks need to be executed in parallel to minimize makespan. Both of these requirements cannot be satisfied due to their complementary nature. Thus the DAG application scheduler needs to take appropriate decision with reference to objective to be achieved. The main challenge while allocating sequential tasks is to minimize communication amount, whereas while allocating parallel tasks it is to minimize execution time.

### B. GridSim and Support for DAG scheduling in GridSim

#### 1) GridSim,

The GridSim is based on SimJava [15]. The SimJava [2] is a discrete event based simulation package [16], which is written in Java language, that provides generalized entities and their behavior that would be needed for any simulator. In this section, we would like to highlight important details of SimJava and GridSim.

In SimJava [15], entities are modeled using Sim_entity class. The Sim_entity provides a body() method which is to be implemented by a derived entity. This body() method can include the logic or activities which get executed as the behavior of the entity when the simulation starts. The SimJava enables interaction among entities through port and event mechanisms. Two entities are connected by input and output ports. The messages are exchanged in form of events. An event is scheduled by using sim_schedule method. The event receiver entity can get events from its associated queues. Two queues are provided to each entity: deferred queue and future event queue. The overall simulation is initiated using Sim_system class. The Sim_system class handles global level activities. The reader can refer [17] for getting more details on SimJava.

The GridSim [9] is a simulation toolkit, not a simulator, to simulate Grid environment and Grid application. The GridSim provides classes for modeling basic entities, such as, Grid resources, jobs, Grid Information Service, etc. The core classes are kept in gridsim package and other additional functionalities are available in sub-packages. The GridSim supports most important functionalities of modeling any real Grid resource's characteristics. The GridSim supports modeling of heterogeneous types of resources. This heterogeneity is possible by modeling resource as time-shared or spaces-shared. The resources can be modeled to have different time-zones and different capabilities. The GridSim has also support for advance reservation of resource, which is useful in performing static scheduling of tasks.

The user entity is responsible for submitting tasks to resources and gathering completed tasks. The GridSim supports regional and global information service, to which various resources registers themselves. The user gets information about registered resources by consulting this Grid Information Service.

### 2) Support for DAG scheduling in GridSim

The original article on GridSim [9] mentions under Application Model that the GridSim has no direct support for scheduling of DAG application. Moreover, we also studied the source code of GridSim version 5.2, a latest version, to find out support for DAG scheduling in it. However, we found that the researchers need to implement such application level functionalities themselves.

Although the GridSim has no direct support for DAG application scheduling, it is possible to extend the GridSim toolkit. The GridSim supports many basic activities such as submitting tasks, receiving completed tasks, querying Grid Information Service, and reserving resources in advance.

### III. DESIGN AND IMPLEMENTATION OF DAG APPLICATION SCHEDULING SIMULATOR

This section discusses architecture of proposed DAG application scheduling simulator. It also discusses about implementation of GUI part and scheduling part.

### A. Design of DAG application scheduling simulator

#### 1) The architecture of the proposed DAG application scheduling simulator

The architecture of the proposed DAG application scheduling simulator is shown in Figure 2. As shown in this diagram, we use SimJava framework for core simulation facilities. For modeling Grid resources, Grid Information Service, and Grid tasks, we use GridSim toolkit. The *Application Scheduler* is in *Application Broker layer*. This Application Scheduler carries out activities that are involved in scheduling tasks of a DAG application. The DAG Application Scheduler GUI, which works as Grid Application layer, has three GUI consoles: DAG GUI, Resource GUI, and Schedule GUI. The DAG GUI is responsible for creating a DAG application by performing drag and drop of tasks and links. The Resource GUI is responsible for taking Grid resource information from the user. The Scheduler GUI is responsible for displaying the assignment of tasks of DAG on the selected resources. The decision regarding which task should be assigned to which resource is taken by the Application Scheduler. The researcher who wants to use this proposed DAG scheduling simulator can change the scheduling algorithm in this Application Scheduler component for study and research.

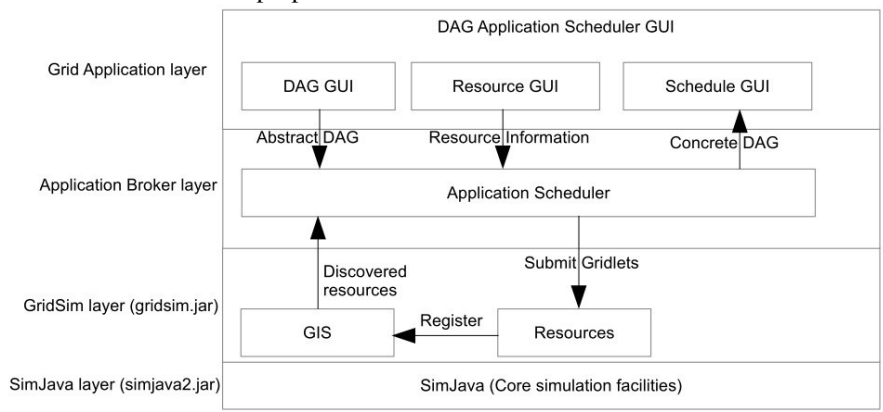

Figure 2 Architecture of proposed DAG application scheduling simulator. This architecture is adapted from higher level architecture of GridSim [9]

#### 2) Features of DAG application scheduling simulator

We list here features of our proposed DAG application scheduling simulator in three categories: (i) task related, (ii) communication link related, and (iii) DAG application related. All these mentioned features are supported by our implemented scheduler.

Task related features:
- Create task using mouse click
- Change position of task by dragging it
- Configure task information through dialog box
- Delete link using mouse click and delete key
- On deleting task, all associated links get deleted automatically

Communication link related features:
- Create link by dragging mouse from a source task and releasing on a destination task
- Configure dependency-link through dialog box
- Delete link using mouse click and delete key

DAG application related features:

It does not accept incorrect DAG application. I.e., it validates DAG.

- A task has no outgoing link
- The DAG application has more than one entry or exit tasks
- The DAG application has floating task (Floating task is a task that is not connected with any other task of the DAG application)

### B. Implementation of DAG application scheduling simulator

The implementation of the proposed DAG scheduling simulator is done using NetBeans IDE 7.0.1, GridSim 5.2 toolkit, and SimJava 2 toolkit. In this section, we discuss important details of the implementation. The implementation is separated in two parts: GUI and scheduling.

The implementation of the proposed DAG scheduling simulator contains many classes. However, we have summarized important classes in Table I. The classes Task, TaskShape, and TaskLink are used in implementation of GUI part, whereas the remaining classes are used in scheduling part.

TABLE I
SOME MAIN CLASSES AND THEIR PURPOSE IN IMPLEMENTATION OF DAG
APPLICATION SCHEDULING SIMULATOR

| Class | Purpose |
|---|---|
| Gridlet | Represents task in Grid environment |
| ARGridResource | Represents resource supporting advance reservation |
| ARSimpleSpaceShared | Represents the resource scheduling policy for ARGridResource |
| AdvanceReservation | Our application scheduler is extended from this class |
| GridInformationService | Provides registry service for resource registration and resource discovery |
| ExtGridlet | This class is extended from the inbuilt Gridlet class. It enables linking of the task with other tasks and task-traversal. |
| ApplicationBroker | This class is extended from AdvanceReservation class and it becomes the user of all the resources supporting ARSimpleSpaceShared resource usage policy. |
| IStaicScheduler | It is an interface having declaration of methods which should be implemented by any static scheduler |
| GridletCollector | This class collects ExtGridlet that finish its execution. |
| Task | This class is used in DAG application GUI. It maintains Information about task. |
| TaskShape | This class is used in DAG application GUI. It is used for display of task on GUI. Methods such as hitTest, draw, highlight methods are implemented in this class |
| TaskLink | This class is used in DAG application GUI. It maintains dependency link between two tasks. |

The main GUI of the DAG application scheduling simulator is shown in Figure 3. The DAG application panel is implemented by extending JPanel class. This panel provides flicker-free GUI, while it needs to be repainted. There are three types of tasks in DAG application, namely Entry-task, Exit-task, and Intermediate task. Each type of task is painted on the DAG panel with different color. The user can select either task or task-link by clicking on its shape. Moreover, the currently selected task or task-link is shown in different color. This Screen-shot shows a DAG application having an intermediate task which has only incoming link and no outgoing link. When, we try to submit such DAG application, it is not taken for scheduling, and user is prompted with appropriate error message, see Figure 3, to correct the DAG. The unit of computation for task is Million Instructions (MI), e.g., task A has 200000 MIPS, and the unit of data-communication is in Bytes, e.g., data-communication between task A and task B is 100000 bytes.

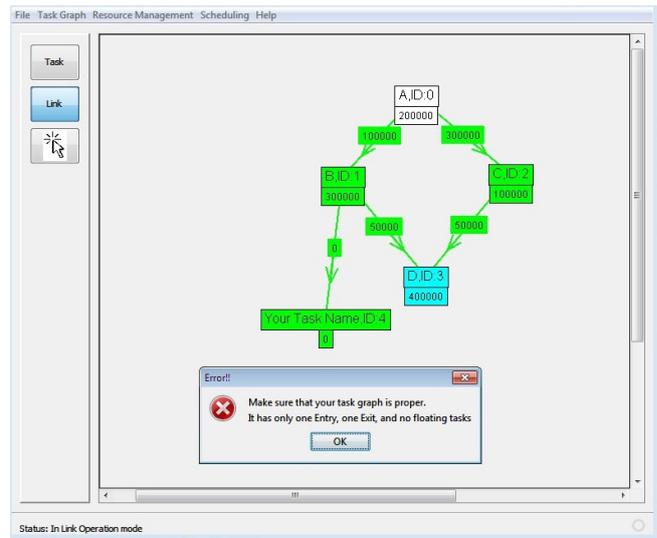

Figure 3 Main GUI of the DAG application scheduling simulator. It validates the DAG before scheduling.

The Figure 4 shows how the Grid resources can be configured for the scheduling. Through this GUI, we can configure resource information such as name of resource, architecture, time-zone, number of machines, number of PEs, baud rate, PE rating, and usage cost of each resource. The user supplies resource information through JTable, see Figure 4. This GUI is dynamic in which we can add or delete row for resource. When this GUI is closed, the contained resource information is retrieved from TableModel. This retrieved resource information is used later for creating Grid resources before simulation is started.

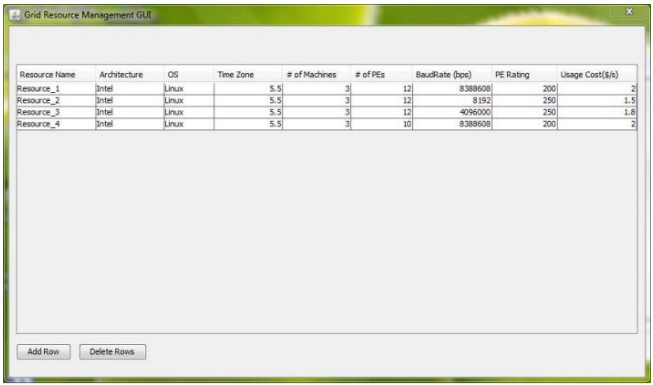

Figure 4 GUI for configuring Grid resources.

Application scheduler is implemented in ApplicationBroker class by extending it from the AdvanceReservation class of GridSim. The behavior of the application scheduler is defined in body() method. The body method first creates an instance of GridletCollector class and an instance of the class that implements IStaticScheduler interface. Then, the ApplicationBroker class invokes the schedule() method on the implementation of IStaticScheduler interface.

Figure 5 shows the code fragment of the GridletCollector class. The GridletCollector class is responsible for collecting all the Gridlets that finish their execution on allocated resources. In the shown code fragment, noOfTasks is the total number of tasks available in the submitted DAG application. The receivedGridletList is the LinkedList, which maintains all collected Gridlets. This receivedGridletList is used by Task-Assignment GUI to prepare the Gannt chart.

```
Sim_event ev=new Sim_event();
Gridlet gridlet = null;

for (int i = 0; i < noOfTasks; i++) {
    gridlet=super.gridletReceive();
    receivedGridletList.add(gridlet);
}
```

Figure 5 Code fragment of GridletCollector class showing how finished Gridlets are collected

## IV. IMPLEMENTATION OF MIN-MIN STATIC SCHEDULING IN PROPOSED DAG SCHEDULING SIMULATOR

Figure 6 shows IStaticScheduler interface. This interface declares methods which are to be implemented by any static task scheduling algorithm. In the schedule method, the scheduling algorithm is implemented. After scheduling certain task or tasks, using getImmdiateUnscheduledTasks() method, next immediate successor tasks can be retrieved. The assignment of task on resource can be recorded using storeTaskAssignment() method. The getScheduleStatus() enables knowing weather all tasks are scheduled or not.

```
public interface IStaticScheduler {
    public void schedule();
    public ExtGridlet[] getImmediateUnscheduledTasks();
```

```
    public void storeTaskAssignment(ExtGridlet
extGridlet,ResourceInformation resourceInformation);
    public boolean getScheduleStatus();
}
```

Figure 6 IStaticScheduler interface

We implement the IStaticScheduler interface in MinMinStaticScheduler. The min-min scheduling algorithm [18], [19] is implemented in schedule() method of MinMinStaticScheduler.java class. The Pseudo code of implemented min-min scheduling algorithm is shown in Figure 7.

```
Perform resource discovery using GIS
Order available resources in order of faster resources first
Initialize #currentlyScheduledTasks=0
Initialize available start-time for each resource
While all tasks are not scheduled
Begin
    ImmUnsTasks = getImmediateUnscheduledTasks()
    For each task (t) of ImmUnsTasks
    Begin
        For each available resource (r)
        Begin
            Calculate data available time
            Calculate execution duration
            Calculate possible reservation start time
            Record possible Completion time based on above values in
                ECT[t][r]
        End
        Record minimum ECT for current task in minECT[t]
    End
    Choose a task that has minimum ECT value in minECT[t].
    Perform advance-reservation for the chosen task
End
Change schedule status from false to true
```

Figure 7 Pseudo code of the min-min static scheduling algorithm implemented in our DAG application scheduling simulator. The above pseudo code provides concise understandingof how it is implemented in MinMinStaticScheduler.java file. The actual lines of code in MinMinStaticScheduler is more than 100.

The above implemented scheduling algorithm is tested for various task graphs having different number of tasks, different resource configuration, and different values of computation time and communication amount. Here, we show result produced for the task graph shown in Figure 8.

The allocation of tasks of the DAG application, shown in Figure 8, is shown in Figure 9. The Task-Assignment GUI shows resource information, task information, and assignment of tasks in form of Gannt chart.

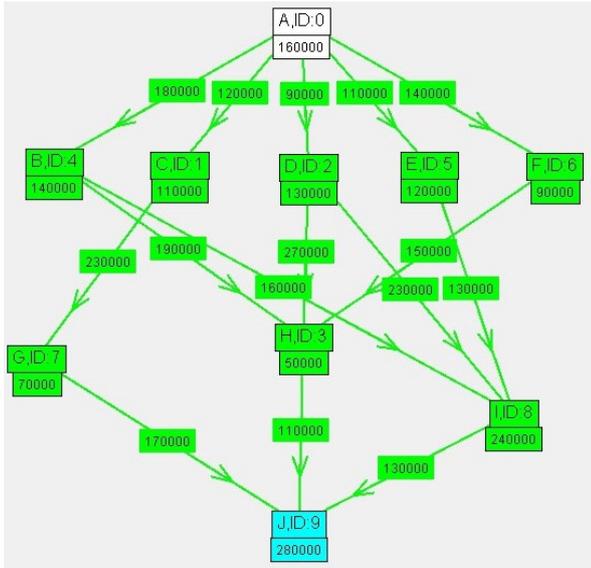

Figure 8 A task-graph representing a DAG application having 10 tasks nodes

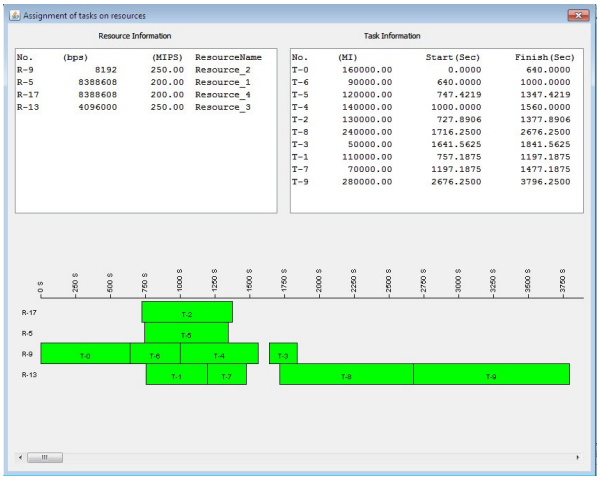

Figure 9 Allocation of tasks of DAG application on Grid resources in Task-Assignment GUI

## V. CONCLUSION

The implemented DAG application scheduling simulator can be used for carrying out performance analysis of DAG/workflow scheduling algorithm. The DAG application scheduling simulator has user interactive GUI for performing activities of configuring DAG and resource information. The DAG application scheduling simulator also shows allocation of tasks of DAG on selected resources in form of Gannt chart in Task-Assignment GUI. The min-min scheduling algorithm is implemented and tested in DAG application scheduling simulator to validate usage of classes involved during scheduling of DAG application.

In future, we would like to work on following further work. At present, in task-graph, computation amount is represented in MI, and communication amount is in Bytes. We would like to allow communication amount in different units and datarate in different units in resource configuration. The default unit of datarate is bps (bits per second). The default tracing provided by the SimJava contains a lot of details. We would like to introduce customized tracing.

## VI. REFERENCES

[1] I. Foster and C. Kesselman, Eds., *The Grid 2: Blueprint for a New Computing Infrastructure*, 2nd ed., ser. The Elsevier Series in Grid Computing. Elsevier, 2003.

[2] P. Brucker, *Scheduling Algorithms*, 3rd ed. Secaucus, NJ, USA: Springer-Verlag New York, Inc., 2001.

[3] O. Sinnen, *Task Scheduling for Parallel Systems (Wiley Series on Parallel and Distributed Computing)*. Wiley-Interscience, 2007.

[4] M. Pinedo, *Scheduling: Theory, Algorithms and Systems*, 2nd ed. Prentice Hall.

[5] M. R. Garey, R. L. Graham, and D. S. Johnson, "Performance guarantees for scheduling algorithms," *Operations Research*, vol. 26, pp. 3–21, 1978.

[6] E. Deelman, G. Singh, M.-H. Su, J. Blythe, Y. Gil, C. Kesselman, G. Mehta, K. Vahi, G. B. Berriman, J. Good, A. Laity, J. C. Jacob, and D. S. Katz, "Pegasus: A framework for mapping complex scientific workflows onto distributed systems," *Sci. Program.*, vol. 13, pp. 219–237, July 2005. [Online]. Available: http://dl.acm.org/citation.cfm?id=1239649.1239653

[7] J. Yu, R. Buyya, and K. Ramamohanarao, "Workflow scheduling algorithms for grid computing," in *Metaheuristics for Scheduling in Distributed Computing Environments*, ser. Studies in Computational Intelligence, F. Xhafa and A. Abraham, Eds. Springer Berlin / Heidelberg, 2008, vol. 146, pp. 173–214, 10.1007/978-3-540-69277-5_7. [Online]. Available: http://dx.doi.org/10.1007/978-3-540-69277-5_7

[8] J. Yu and R. Buyya, "A budget constrained scheduling of workflow applications on utility grids using genetic algorithms," in *Workshop on Workflows in Support of Large-Scale Science, Proceedings of the 15th IEEE International Symposium on High Performance Distributed Computing (HPDC)*. IEEE, IEEE CS Press, 2006.

[9] R. Buyya and M. Murshed, "Gridsim: a toolkit for the modeling and simulation of distributed resource management and scheduling for grid computing," *Concurrency and Computation: Practice and Experience*, vol. 14, no. 13-15, pp. 1175–1220, 2002. [Online]. Available: http://dx.doi.org/10.1002/cpe.710

[10] H. Casanova, "Simgrid: a toolkit for the simulation of application scheduling," in *Cluster Computing and the Grid, 2001. Proceedings. First IEEE/ACM International Symposium on*, 2001, pp. 430 –437.

[11] "Chicsim (the chicago grid simulator)," http://people.cs.uchicago.edu/ krangana/ChicSim.html. [Online]. Available: http://people.cs.uchicago.edu/~krangana/ChicSim.html

[12] W. H. Bell, D. G. Cameron, L. Capozza, A. P. Millar, K. Stockinger, and F. Zini, "Optorsim - a grid simulator for studying dynamic data replication strategies," *International Journal of High Performance Computing Applications*, vol. 17, no. 4, pp. 403–416, 2003.

[13] D. Kondo, "SimBOINC: A simulator for desktop grids and volunteer computing systems." [Online]. Available: http://simboinc.gforge.inria.fr/

[14] D. Klusáček and H. Rudová, "Alea 2: job scheduling simulator," in *Proceedings of the 3rd International ICST Conference on Simulation Tools and Techniques*, ser. SIMUTools '10. ICST, Brussels, Belgium, Belgium: ICST (Institute for Computer Sciences, Social-Informatics and Telecommunications Engineering), 2010, pp. 61:1–61:10. [Online]. Available: http://dx.doi.org/10.4108/ICST.SIMUTOOLS2010.8722

[15] F. Howell and R. McNab, "Simjava: A discrete event simulation library for java," *Simulation Series*, vol. 30, pp. 51–56, 1998.

[16] R. Jain, *The art of computer systems performance analysis*. John Wiley & Sons, 2008.

[17] C. Simatos, "The simjava tutorial," *University of Edinburgh,(http://www. icsa. inf. ed. ac. uk/research/groups/hase/simjava/guide/tutorial. html email: C. Simatos@ sms. ed. ac. uk)*, 2002.

[18] J. Blythe, S. Jain, E. Deelman, Y. Gil, K. Vahi, A. Mandal, and K. Kennedy, "Task scheduling strategies for workflow-based

applications in grids," in *Cluster Computing and the Grid, 2005. CCGrid 2005. IEEE International Symposium on*, vol. 2. IEEE, 2005, pp. 759–767.

[19] X. He, X. Sun, and G. Von Laszewski, "Qos guided min-min heuristic for grid task scheduling," *Journal of Computer Science and Technology*, vol. 18, no. 4, pp. 442–451, 2003.